\documentclass[prd,reprint,showkeys]{revtex4}%
\usepackage{amssymb}
\usepackage{amsfonts}
\usepackage{amsmath}
\usepackage{graphicx}%
\providecommand{\U}[1]{\protect\rule{.1in}{.1in}}

\begin{document}
\title{Hawking Radiation of Mass Generating Particles From Dyonic Reissner
Nordstr\"{o}m Black Hole }
\author{I. Sakalli}
\email{izzet.sakalli@emu.edu.tr}
\author{A. \"{O}vg\"{u}n}
\email{ali.ovgun@emu.edu.tr}
\affiliation{Physics Department, Eastern Mediterranean University, Famagusta, Northern
Cyprus, Mersin 10, Turkey}
\date{\today }

\begin{abstract}
The Hawking radiation is considered as a quantum tunneling process, which can
be studied in the framework of the Hamilton-Jacobi method. In this study, we
present the wave equation for a mass generating massive and charged scalar
particle (boson). In sequel, we analyze the quantum tunneling of these bosons
from a generic 4-dimensional spherically symmetric black hole. We apply the
Hamilton-Jacobi formalism to derive the radial integral solution for the
classically forbidden action which leads to the tunneling probability. To
support our arguments, we take the dyonic Reissner-Nordstr\"{o}m black hole as
a test background. Comparing the tunneling probability obtained with the
Boltzmann formula, we succeed to read the standard Hawking temperature of the
dyonic Reissner-Nordstr\"{o}m black hole.

\end{abstract}
\maketitle

\section{Introduction}

In 1975, Stephen Hawking (one of the world's most famous physicists) made a
shocking claim that when Quantum Mechanics is allied with the General
Relativity, black holes (BHs) become to glow with Hawking radiation (HR)
\cite{Haw1,Haw2,Haw3,Haw4}. This emission consists of all sorts of
massless/massive particles with different spins: spin$-0,1/2,1$.... Hawking's
prodigious calculations are based on a scenario that ubiquitous virtual
particle pairs are continually being created near the event horizon of the BH
due to the vacuum fluctuations. Principally, these particles are created as a
particle-antiparticle pair and immediately after they quickly annihilate each
other. However, it is always possible that the one with negative energy (in
order to conserve the total energy) falls into the BH while the other
possessing the positive energy escapes to spatial energy as HR. Today, HR is
also called the Bekenstein-Hawking radiation in virtue of Bekenstein's
remarkable contributions \cite{bek1,bek2,bek3,bek4} to this phenomenon.

Since 1975, the studies concerning HR have been carrying on. Up to the present
times, many different methods for the HR are proposed (the reader may refer to
\cite{gib1,gib2,vanzo,umetso,kraus1,kraus2} and references therein). Among
them, the most fascinating quantum tunneling methods are Parikh and Wilczek's
null-geodesic method \cite{wilczek1,wilczek2,wilczek3} and the semiclassical
methods of Hamilton-Jacobi \cite{ang,sri1,sri2,man0} and Damour-Ruffini
\cite{ruf}. On the other hand, the HR of photons, scalar particles, massive
vector bosons, and fermions from various BHs have been gained much attention
in recent years (see for example
\cite{sakhal,sak1,Ma,sak2,Ji,sak3,sak4,ao1,mann1,mann2,mann3,xiang,kruglov1,kruglov2,ao2,aoo2,ao3,ao4,ao6,frasca,valt,rocha,gos,sing,zhi,yangg}%
). Furthermore, the information loss paradox \cite{loss1,loss2,loss3} in the
HR is one of the great puzzles for the physics community. Some theorists bring
forward an idea to retrieve the information from the BH encoded in the HR
\cite{haw15,hooft,hooft1,dvali,lochan,stoica,kraus,louko,rman,calmet,perez,papa,shi,marolf,malda}%
. However, this mystery have not been solved literally.

In the 1970s,\ particle physicists realized that there are very close ties
between two of the four fundamental forces
\cite{force,sm1,sm2,sm3,sm4,sm5,sm6} -- the weak force and the electromagnetic
force which is single underlying force known as the electroweak force. The
basic equations of the unified theory correctly describe the relationship
between the electroweak force and its associated force-carrying particles
[photons and the massive vector bosons ($W^{\pm}$ and $Z$)], except for a
major glitch: all of these particles emerge without a mass! Although this is
true for the photon, we know that the $W^{\pm}$ and $Z$ bosons must have mass,
nearly hundred times that of a proton. The problem of spontaneously broken
gauge theories in curved spacetime is well known in literature
\cite{barroso,tro,rand,arkani,witt,dad0,dad1}. So far, the Higgs mechanism
\cite{sm4,sm5} is the experimentally confirmed mechanism to solve the
generation of mass problem in particle physics, which satisfies both the
unitarity and the renormalization of the theory.

In this paper, we make a brief review for the derivation of the wave equation
for the mass generating (massive and charged) scalar particles. Applying the
resulting equation obtained to the general 4-dimensional static and
spherically symmetric metric, we obtain the general radial integral solution
for the action of Hamilton-Jacobi method. As a test bed we consider the dyonic
Reissner-Nordstr\"{o}m BH (DRNBH) \cite{dy} and compute its quantum tunneling
rate by using the latter radial integral solution of the action. Finally, we
show in detail how one recovers the original HR of the DRNBH from the quantum
tunneling of the mass generating particles.

The paper is organized as follows. In Sec. II, we introduce the wave equation
of a massive and charged mass generating scalar particle in a curved
spacetime. Section III is devoted to the computations of the quantum tunneling
of the mass generating scalar particles from the DRNBH. While doing this, we
are attentive to make our calculations with generic as much as possible. We
draw our conclusions in Sec. IV.

\section{Wave Equation of Mass Generating Particles}

In this section, we represent an expression for the wave equation of the mass
generating particles. Their associated scalar fields are non-minimally coupled
to the gravity. The main idea underlying this mass generation mechanism is
resplendently introduced in many textbooks (see for instance \cite{qft1,la}).

For brevity, we initially use units $G_{N}=c=\hbar=1$. One may write down the
action of the interaction of the scalar fields with gravity \cite{barroso} as follows%

\begin{equation}
S=\int d^{4}x\sqrt{-g}\left[  \frac{\Re}{16\pi}-\xi\phi^{\dag}\phi\Re+(D_{\mu
}\phi)^{\dag}D^{\mu}\phi-V(\phi)-\frac{1}{4}F_{\mu\nu}F^{\mu\nu}\right]  ,
\label{1}%
\end{equation}

where $\Re$ stands for the scalar curvature and $F_{\mu\nu}=\nabla_{\mu}%
A_{\nu}-\nabla_{\nu}A_{\mu}$ is the Maxwell field strength with the spin-1
gauge field $A_{\nu}$ (electromagnetic vector potential). $\xi$ denotes the
dimensionless coupling constant which governs the non-minimal interaction of
the scalar field $\phi$ \cite{foot1} with gravity. In other words, the
minimally coupled scalar fields correspond to $\xi=0$. It is worth noting that
this coupling constant $\xi$ can also be used to stabilize the vacuum
expectation value $y^{2}=\frac{v^{2}}{2}=\left\langle \phi^{\dag}%
\phi\right\rangle $ near the event horizon of a BH \cite{dad1}. The
gauge-covariant derivative is given by
\begin{equation}
D_{\mu}=\partial_{\mu}-ieA_{\mu}, \label{2}%
\end{equation}
where $e$ is the coupling constant (i.e. the Planck charge) of the
electromagnetic vector potential $A_{\mu}$. The variation of the action (1)
with respect to the metric tensor $g_{\mu\nu}$ leads to the Einstein equations
of motion as follows%

\begin{equation}
\Re_{\mu\nu}-\frac{1}{2}\Re g_{\mu\nu}=-8\pi T_{\mu\nu}, \label{3}%
\end{equation}

where $T_{\mu\nu}$ is the energy-momentum tensor. Its overlong expression can
be seen in the study of Moniz et al \cite{barroso}. Significantly, when one
applies the variation to the action (1) with respect to $\phi^{\dag}$, the
following wave equation is obtained%

\begin{equation}
\frac{1}{\sqrt{-g}}\left(  \partial_{\mu}-ieA_{\mu}\right)  \left[  \sqrt
{-g}g^{\mu\nu}\left(  \partial_{\nu}-ieA_{\nu}\right)  \phi\right]  +\xi
\Re\phi+\partial_{\phi^{\dag}}V=0. \label{4}%
\end{equation}

The mass generating potential was also defined in \cite{barroso} as follows%

\begin{equation}
V(\phi)=B+\widetilde{m}^{2}\phi^{\dag}\phi+\lambda(\phi^{\dag}\phi)^{2},
\label{5}%
\end{equation}

where $B$ is an arbitrary constant and the coupling constant $\lambda$ is
dimensionless in 4-dimensional spacetime. Without loss of generality, it is
assumed that $\lambda$ has a positive definite value. As clearly stated in
\cite{barroso}, the vacuum expectation value must satisfy the condition of
$y^{2}\neq0$, which requires that $V(\phi)$ must have a minimum at $\phi\neq
0$. To obtain the bounded solution for the Hamiltonian, $\widetilde{m}^{2}$
must be negative since $\lambda$ is positive. Due to this reason, we shift
$\widetilde{m}^{2}\rightarrow-m^{2}$. Hence, using Eq. (5), we have%

\begin{equation}
\partial_{\phi^{\dag}}V=\left[  -m^{2}+2\lambda(\phi^{\dag}\phi)\right]  \phi.
\label{6}%
\end{equation}

After assigning the reduced Planck constant back to its original value
$\hslash$, Eq. (4) can be rewritten as%

\begin{equation}
\frac{1}{\sqrt{-g}}\left(  \partial_{\mu}-i\frac{e}{\hslash}A_{\mu}\right)
\left[  \sqrt{-g}g^{\mu\nu}\left(  \partial_{\nu}-i\frac{e}{\hslash}A_{\nu
}\right)  \phi\right]  +\frac{1}{\hslash^{2}}\left[  \xi\Re-m^{2}%
+2\lambda(\phi^{\dag}\phi)\right]  \phi=0. \label{7}%
\end{equation}

which is the wave equation of the mass generating particles with mass $m$ and
charge $e$ in a curved spacetime. It is also important to know that whenever
the scalar field $\phi$ is used for a Nambu--Goldstone boson in the gauge
theory of spontaneous symmetry breaking, $\xi$ is zero \cite{volo}. On the
other hand, if the scalar field $\phi$ represents a composite particle, then
the value of $\xi$ is fixed by the dynamics of its components. In particular,
$\xi=1/6$ in the large $N$ approximation to the Nambu-Jona-Lasinio model
\cite{hill}. Moreover, in the standard model, the Higgs fields possess the
values of $\xi$ within the range of $\xi\leq0$ and $\xi\geq1/6$ \cite{hoso}.

\section{Quantum Tunneling of Mass Generating Particles from DRNBH}

The line-element for the 4-dimensional generic static (spherically symmetric)
BH metric is given by%
\begin{equation}
ds^{2}=-Fdt^{2}+G^{-1}dr^{2}+R\left(  d\theta^{2}+\sin^{2}\theta d\varphi
^{2}\right)  , \label{8}%
\end{equation}
where the metric functions $\left(  F,G,R\right)  $ are only the function of
$r$. Any horizon $r_{h}$ should satisfy the condition of $G(r_{h})=0$ and
$r_{h}$ is, in general, a function of the mass and charge of the BH. The
Hawking temperature of a BH described by the metric (8) is given by \cite{fer}%

\begin{equation}
T_{H}=\frac{1}{4\pi}\left.  \left\vert \frac{dg_{tt}}{dr}\right\vert
\sqrt{-g^{tt}g^{rr}}\right\vert _{r=r_{h}}=\frac{F^{\prime}\left(
r_{h}\right)  }{4\pi\sqrt{B\left(  r_{h}\right)  }}, \label{9}%
\end{equation}

where $B=\frac{F}{G}$ and the prime over a quantity denotes the derivative
with respect to $r$. Furthermore, the Ricci scalar \cite{wald} for the metric
(8) can be found as%

\begin{align}
\Re &  =\frac{1}{2F^{2}R^{2}}(-G^{\prime}F^{\prime}FR^{2}-2F^{\prime\prime
}FGR^{2}+F^{^{\prime}2}GR^{2}-\nonumber\\
&  2RGF^{\prime}R^{\prime}F+R^{\prime2}F^{2}G-4R^{\prime\prime}F^{2}%
GR-2G^{\prime}R^{\prime}F^{2}R+4F^{2}R). \label{10}%
\end{align}

In order to study the quantum tunneling of the mass generating particles from
the generic BH (8), we use the WKB approximation and assume an ansatz for the
scalar field $\phi$ as follows
\begin{equation}
\phi=c\exp\left(  \frac{i}{\hbar}I(t,r,\theta,\varphi)\right)  , \label{11}%
\end{equation}

where $c$ is the amplitude of the wave and $I$ stands for the classically
forbidden action of the trajectory. Metric (8) admits two Killing vectors
$<$
$\partial_{t}$, $\partial_{\varphi}$
$>$%
, which show the existence of the symmetries. Therefore, one can assume a
solution for the action as \
\begin{equation}
I=-Et+W(r)+j(\theta,\varphi)+C, \label{12}%
\end{equation}

where $E$ denotes energy, $W(r)$ and $j(\theta,\varphi)$ are radial and
angular functions, respectively. In Eq. (12) $C$\ is a complex constant.

Since $A_{\nu}$ represents the electromagnetic vector potential, for a dyonic
BH with electric and magnetic components one should have $A_{\nu}=\left[
A_{0}(r),0,0,A_{1}(\theta)\right]  $. Under the guidance of the
Hamilton-Jacobi method \cite{ang}, we first insert Eqs. [11-13] in Eq. (7) and
then consider the terms with the leading order of $\hbar$. Thus, we obtain the
following lenghty expression%

\begin{equation}
\sin^{2}\theta\{\left[  (-G^{\prime}F^{\prime}-2F^{\prime\prime}%
G)F+F^{\prime2}G\right]  R^{2}-2\left[  \left(  G^{\prime}R^{\prime
}-2+2GR^{\prime\prime}\right)  F+GF^{\prime}R^{\prime}\right]  FR+R^{\prime
}F^{2}G\}\xi\nonumber
\end{equation}

\begin{equation}
-2FR\left\{  \left[  (-m^{2}-2\,\lambda c^{2}-GW^{\prime2})R-{j_{\theta}}%
^{2}\right]  \mathrm{\sin}^{\mathrm{2}}\theta-(eA_{1}-{j_{\varphi}}%
)^{2})F+R\mathrm{\sin}^{\mathrm{2}}\theta E_{net}^{2}\right\}  =0, \label{13}%
\end{equation}

where ${j_{\theta}}=\frac{\partial j}{\partial\theta}$, ${j_{\varphi}}%
=\frac{\partial j}{\partial\varphi}$, and $E_{net}=E+eA_{0}$. From Eq. (13),
we derive an integral solution for $W(r)$ as follows%

\begin{equation}
W_{\pm}=\pm\int\frac{1}{\sqrt{FG}}\left[  \left(  n_{1}F+n_{2}\right)  \xi
G+\frac{F}{R}n_{3}+E_{net}^{2}\right]  ^{\frac{1}{2}}dr, \label{14}%
\end{equation}

where%

\begin{equation}
n_{1}={\frac{2R^{\prime\prime}}{R}}+{\frac{G^{\prime}R^{\prime}-2}{GR}}%
-{\frac{1}{2}}\left(  {\frac{R^{\prime}}{R}}\right)  ^{2}, \label{15}%
\end{equation}

\begin{equation}
n_{2}=\frac{F^{\prime2}}{2F}-\frac{F^{\prime}R^{\prime}}{R}-\frac{1}{2}%
\frac{G^{\prime}F^{\prime}}{G}-F^{\prime\prime}, \label{16}%
\end{equation}

\begin{equation}
n_{3}=(-2c^{2}\lambda-m^{2})R-{j_{\theta}}^{2}-\frac{(eA_{1}-{j_{\varphi}%
})^{2}}{\mathrm{\sin}^{\mathrm{2}}\theta}. \label{17}%
\end{equation}

Since $G(r_{h_{+}})=0$, the near horizon form of Eq. (14) becomes%

\begin{equation}
W_{\pm}\approx\pm\int\sqrt{\frac{E_{net}^{2}R+Fn_{3}}{RFG}}dr. \label{18}%
\end{equation}

which is essential expression for computing the quantum tunneling rate. Now,
we test the above result obtained via the DRNBH geometry \cite{dy} whose the
metric functions and electromagnetic vector potential components are given by%

\begin{equation}
F=G=\frac{\left(  r-r_{h_{+}}\right)  \left(  r-rh_{_{-}}\right)  }{r^{2}},
\label{19}%
\end{equation}

\begin{equation}
R=r^{2}, \label{20}%
\end{equation}

\begin{equation}
A_{0}=-\frac{Q}{r},\text{ \ \ }A_{1}=P\cos\theta. \label{21}%
\end{equation}

where the physical quantities $Q$ and $P$ denote the DRNBH's characteristic
parameters: $Q$ is the electric charge and $P$ is the magnetic charge. The
outer or event ($r_{h+}$) and inner ($r_{h-}$) horizons of the DRNBH are given by%

\begin{equation}
r_{h_{\pm}}=M\pm\sqrt{M^{2}-\Theta^{2}}, \label{22}%
\end{equation}

where $\Theta^{2}=Q^{2}+P^{2}$. In Eq. (22) the parameter $M$ represents the
mass of the DRNBH. Since $F=G,$ consequently $Fn_{3}\rightarrow0$ around the
event horizon. Here, one can immediately criticize why the $n_{3}$ parameter
including the particle's mass $m$\ quickly drops out of the considerations.
However, one can experience from the previous studies \cite{vanzo} that the
non-differential terms coupled to the wave function $\phi$\ (for example, in
Eq. (7), it corresponds to $\frac{1}{\hslash^{2}}\left[  \xi\Re-m^{2}%
+2\lambda(\phi^{\dag}\phi)\right]  \phi$ ) apart from the operator term acting
on $\phi$ (like the Laplacian operator: $\square\phi$) always looses its
effiency near the horizon. That is why, for instance, the HR is independent
from the particle's mass \cite{ex1,ex2}. Thus, Eq. (18) reduces to%

\begin{equation}
W_{\pm}\approx\pm\int\frac{E_{net}}{F}dr. \label{23}%
\end{equation}

Meanwhile, we now have $E_{net}=E-\frac{eQ}{r_{h_{+}}}$. It is obvious that
the above integrand possesses a simple pole at the event horizon. To evaluate
integral (23), we first expand the metric function $F$ as follows%

\begin{equation}
F(r)=F^{\prime}(r_{h_{+}})(r-r_{h_{+}})+\Game(r-r_{h_{+}})^{2}. \label{24}%
\end{equation}

Substituting the above expression into Eq. (23) and choosing the contour as a
half loop going around this pole from left to right, one obtains%

\begin{equation}
W_{\pm}=\pm i\pi\frac{E_{net}}{F^{\prime}\left(  r_{h_{+}}\right)  }.
\label{25}%
\end{equation}

Thus, the imaginary part of the action (12) becomes%

\begin{equation}
\operatorname{Im}I_{\pm}=ImC\pm\pi\frac{E_{net}}{F^{\prime}\left(  r_{h_{+}%
}\right)  }. \label{26}%
\end{equation}

Thence, we compute the probabilities of ingoing and outgoing particles
tunneling the DRNBH horizon as%

\begin{equation}
P_{in}=\exp(-2ImI_{-})=\exp\left(  -2ImC+2\pi\frac{E_{net}}{F^{\prime}\left(
r_{h_{+}}\right)  }\right)  , \label{27}%
\end{equation}

\begin{equation}
P_{out}=\exp(-2ImI_{+})=\exp\left(  -2ImC-2\pi\frac{E_{net}}{F^{\prime}\left(
r_{h_{+}}\right)  }\right)  . \label{28}%
\end{equation}

Classically, having a BH is conditional on the no-reflection for the ingoing
waves, which meants full absorption: $P_{in}=1$. This is possible simply by
setting $ImC=\pi\frac{E_{net}}{F^{\prime}\left(  r_{h_{+}}\right)  }$ (for
similar and recent works, the reader is referred to \cite{con1,con2,con3} and
references therein) which results in%

\begin{equation}
P_{out}=\exp\left(  -4\pi\frac{E_{net}}{F^{\prime}\left(  r_{h_{+}}\right)
}\right)  . \label{29}%
\end{equation}

Consequently, we read the quantum tunneling rate for the DRNBH as%

\begin{equation}
\Gamma=\frac{P_{out}}{P_{in}}=\exp\left(  -4\pi\frac{E_{net}}{F^{\prime
}\left(  r_{h_{+}}\right)  }\right)  . \label{30}%
\end{equation}

Employing the Boltzmann formula $\Gamma=\exp(-E_{net}/T)$ \cite{bolt}, the
surface temperature of the DRNBH can be computed as%

\begin{equation}
T=\frac{F^{\prime}\left(  r_{h_{+}}\right)  }{4\pi}=\frac{r_{h_{+}}-r_{h_{-}}%
}{r_{h_{+}}^{2}}=\frac{\sqrt{M^{2}-\Theta^{2}}}{2\pi\left(  M+\sqrt
{M^{2}-\Theta^{2}}\right)  ^{2}}, \label{31}%
\end{equation}

which is exactly equal to the standard Hawking temperature of the DRNBH
\cite{dy}. Temperature versus mass plotting is depicted in Fig. (1) for
$M\geq\Theta$. As it can be seen from Fig. (1), the locations of the peaks on
the $M$-axis (which are very close to their associated starting mass value
$M_{initial}=\Theta$: the extreme BH case, $T=0$) shift towards right with increasing
$M$-value, however the peak values decrease when $M_{initial}$ gets bigger numbers.
Moreover, while $M\rightarrow\infty,$ all the curves of the temperatures
rapidly reach to the curve of the Schwarzschild ($\Theta=0$) BH's Hawking
temperature, which goes to zero with increasing $M$-value. 

\begin{figure}[h]
\centering
\includegraphics[scale=.75]{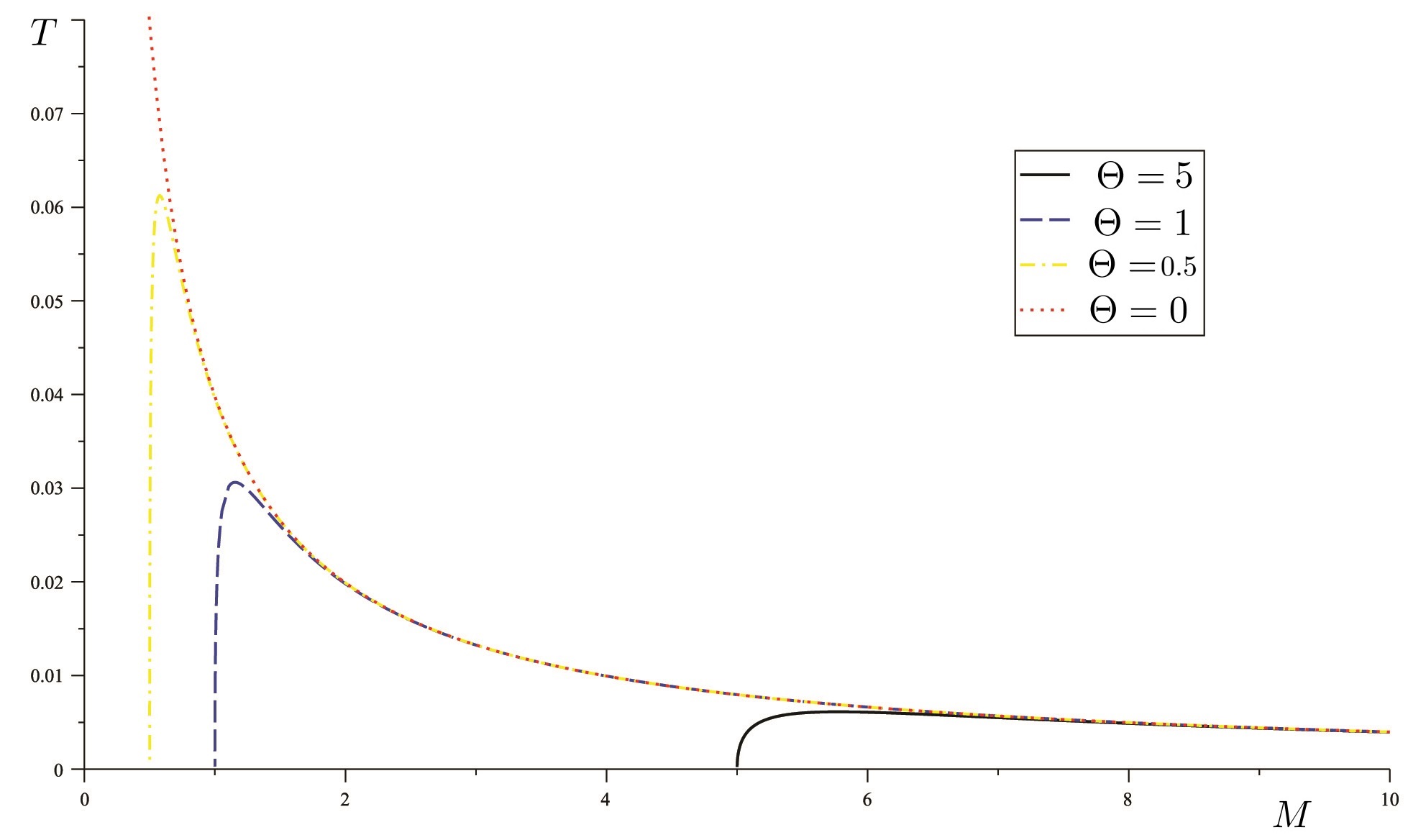}\caption{Plots of the temperature
$T$ versus DRNBH mass $M$. The plots are governed by Eq. (31). The starting
masses are governed by $M_{initial}=\Theta$.}%
\end{figure}

\section{Conclusion}

In this paper, we firstly reviewed the derivation of the wave equation for the
mass generating scalar particles in the concept of the spontaneous symmetry
breaking theory. To this end, we introduced an action involving a non-minimal
scalar field coupled to the gravity. By using the Hamilton-Jacobi method with
a suitable WKB ansatz, the quantum tunneling of the mass generating bosons
from a generic static BH is thoroughly studied. We then obtained the general
integral solution for the radial function (14) for the Hamilton-Jacobi action
$I$. DRNBH geometry whose the metric functions satisfy the equality $F=G$ is
considered as a test background for our computations. It is seen that scalar
particle mass $m$, the non-minimal coupling constant $\xi$, and the potential
constant $\lambda$ are not decisive for the quantum tunneling rate, however
the charge $e$ is. In the semiclassical framework, we computed the
probabilities of the ingoing and outgoing particles to get the quantum
tunneling rate for the DRNBH. Finally, we managed to read the standard Hawking
temperature of the DRNBH via the Boltzmann formula of the tunneling rate.

In future work, we plan to extend our analysis to a BH (might be a spherically
non-symmetric) having $F$, which does not vanish at the event horizon:
$F(r_{h})\neq0$. Because in such a case Eq. (18) may yield such $W_{\pm}$
values (having now the potential constant term $\lambda$ ) that the quantum
tunneling rate can deviate from its pure thermal character \cite{foot2} and
give contribution to the information loss problem \cite{loss3}. We also aim to
extend our analysis to the dynamic, rotating and higher/lower dimensional BHs.
In this way, we will analyze the HR of the mass generating particles from
various BHs.

\section*{Acknowledgements}

The authors are grateful to the Editor and anonymous Referees for their
valuable comments and suggestions to improve the paper.


\begin{thebibliography}{99}
\bibitem {foot2}Thermal radiations do not carry information, see for example
\cite{wilczek1}.
\end{thebibliography}
\end{document}